\documentclass[11pt,letterpaper,twocolumn]{article}
\usepackage{amsthm}
\usepackage{amssymb}
\usepackage{amsmath}
\usepackage{latexsym}
\usepackage[dvips]{graphicx}
\usepackage{abstract}
\newtheorem{thm}{Theorem}
\newtheorem{prop}[thm]{Proposition}
\newtheorem{ex}[thm]{Example}

\DeclareMathOperator*{\argmax}{argmax}

\title{Alignment Metric Accuracy}
\author{Ariel S Schwartz \qquad\qquad\qquad
  Eugene W Myers \\\\
  Computer Science Division \\University of California Berkeley\\
  \texttt{\{sariel,gene\}@cs.berkeley.edu}
  \and Lior Pachter\thanks{Corresponding author} \\\\
  Department of Mathematics  \\University of California Berkeley\\
  \texttt{lpachter@math.berkeley.edu} }

\begin{document}
\twocolumn[
\maketitle
\begin{onecolabstract}
We propose a metric for the space of multiple sequence alignments that
can be used to compare two alignments to each other. In the case
where one of the alignments is a reference alignment,
the resulting accuracy measure improves upon
previous approaches, and provides a balanced assessment of the fidelity
of both matches and gaps. Furthermore, 
in the case where a reference alignment is not available, 
we provide empirical evidence 
that the distance from an alignment produced by one program
to predicted alignments from other programs can be used as a control
for multiple alignment experiments. In particular, we show that
low accuracy alignments can be effectively identified and discarded.

We also show that in the case of pairwise sequence alignment, 
it is possible to find an alignment that
maximizes the expected value of our accuracy measure. Unlike previous
approaches based on expected accuracy alignment that tend to  maximize
sensitivity at the expense of specificity, our method is able to
identify unalignable sequence, thereby increasing overall accuracy.
In addition, the algorithm allows for control of the
sensitivity/specificity tradeoff via the adjustment of a single parameter. 
These results are confirmed with simulation studies that show that unalignable
regions can be distinguished from homologous, conserved sequences.

Finally, we propose an extension of the pairwise alignment method to multiple alignment. 
Our method, which we call AMAP, outperforms existing protein sequence multiple 
alignment programs on benchmark datasets. A webserver and software downloads are available at
\newline
{\tt http://bio.math.berkeley.edu/amap/}.
\end{onecolabstract}
]
\saythanks

\section{Introduction}
A recent survey on sequence alignment \cite{batzoglou05} discusses a
number of important problems and challenges that need to be overcome
in order to facilitate large-scale comparative analysis of the
multiple genomes currently being sequenced. Among these, the following
two problems are highlighted:
\begin{enumerate}
\item ``As suggested \cite{miller00}, methods to evaluate alignment accuracy. This
  goes at the core of the problem: which regions are alignable, and
  what is a correct alignment?''
\item ``A definition of {\em alignability} -- at what point is it no
  longer possible to do meaningful sequence alignment. Or rather, at
  what point can one conclude that two sequences are no longer related?''
\end{enumerate}
Furthermore, the development of ``rigorous methods for evaluating the
accuracy of an alignment'' and the need for
``improved pairwise alignment with a statistical basis'' are singled
out as the most pressing challenges for the alignment community. 

Two commonly used alignment accuracy measures are the \emph{developer} ($f_D$)
and the \emph{modeler} ($f_M$) measures \cite{sauder00}. These measures
correspond to evaluating the sensitivity (number of correctly matched pairs 
divided by the number of matched pairs in the reference alignment)  and
specificity (number of correctly matched pairs divided by the number of matched pairs
in the predicted alignment) of matched
pairs in the predicted alignment respectively. 
Both of these measures have a problem, in that they do not account for gap columns. 
\cite{blanchette04} defined the \emph{agreement} score, as the fraction of
pairwise columns in the predicted alignment that agree with the reference alignment.
While the agreement score does consider gap columns, it is not symmetric,
since the number of columns in the predicted alignment can differ from the
number of columns in the reference alignment. 

What properties should an alignment accuracy measure satisfy? If we
think of accuracy measure as a ``distance'', then if $h^i$
and $h^j$ are two alignments and we denote their distance by
$d(h^i,h^j)$, we would like to have:
\begin{eqnarray}
d(h^i,h^j) & \geq & 0, \nonumber\\
       & = & 0 \textrm{ if, and only if, } h^i \sim h^j,\\
d(h^i,h^j) & = & d(h^j,h^i),\\
d(h^i,h^k) & \leq & d(h^i,h^j) + d(h^j,h^k).
\end{eqnarray}
The first condition specifies that the distance between two alignments
should be non-negative and should be $0$ if, and only if, the two
alignments are equivalent (a useful definition of equivalence is given
in section \ref{sec:metric}). The second requirement specifies that the
distance should be symmetric. For example, comparing a prediction with a
reference alignment should be the same as comparing the reference
alignment to the prediction. The third requirement ensures a certain
consistency: the distance between two predictions should be less than
the sum of the distances from the predictions to a reference
alignment (the triangle inequality). In other words, 
an accuracy measure should be based on a {\em
metric}. Furthermore, the accuracy measure should account for unalignable
sequence. For example, if two sequences are unrelated, the true
alignment contains only gaps (regardless of order), and a good
accuracy measure should reflect that. 
Note that although metrics on the space of \emph{sequences} have been constructed \cite{spiro04}, 
a metric for alignment accuracy should be defined on the space of \emph{sequence alignments},
and should measure the distance between alignments not sequences.

In section \ref{sec:metric} we define an alignment metric, 
and a new accuracy measure, called \emph{AMA}, based on this metric.
Next, we propose an algorithm which maximizes the expected AMA
value given an underlying probabilistic model for mutation of
sequences. We term this alignment strategy AMAP. The algorithm is
explained in detail in section \ref{sec:meama}, where we show that a
single parameter we term \emph{gap-factor} ($G_f$) can be used to
adjust the sensitivity/specificity tradeoff. A special case of AMAP,
where we set $G_f = 0$, is the Maximum Expected Accuracy (MEA)
alignment algorithm introduced in \cite{durbin98,holmes98} and
used in ProbCons \cite{do05}, and Pecan \cite{pecan}. 

In section \ref{sec:results} we show how existing algorithms perform when
judged using AMA, and contrast this with the developer
score, which is the standard measure used in most papers 
\cite{do05,edgar04,sze05}.
Since the developer score is a measurement of sensitivity of aligned pairs, and
algorithms have traditionally been judged by it, we find that existing
algorithms are heavily biased in favor of sensitive, rather than
specific alignments. An extreme case of this can be seen in Table \ref{tbl:sim}
where we show that existing algorithms align large
fractions of completely unrelated sequences. We also see that
multiple alignments produced by different programs differ considerably
from each other, even though they may appear to perform similarly when 
judged only by the developer score.

In a different application of the alignment metric, 
we show that in the typical case where reference alignments
are not available for judging the success of multiple alignment
experiments, the metric can be used
as a control by measuring the distances between alignments predicted
by different programs. 

Finally, we analyze the performance of the new AMAP algorithm, 
using the SABmark dataset \cite{vanWalle05} and simulated datasets,
and show that AMAP compares favorably to other programs.

\section{Metric Based Alignment \\Accuracy}
\label{sec:metric}
Following the notation in \cite{pachter05}, an alignment of a pair of sequences
$\sigma^1=\sigma^1_1\sigma^1_2\cdots\sigma^1_n$ and $\sigma^2=\sigma^2_1\sigma^2_2\cdots\sigma^2_m$
can be represented by an \emph{edit string} $h$ over the \emph{edit alphabet} \{$H,I,D$\},
where $H$ stands for homology, $I$ for insertion, and $D$ for deletion.
Equivalently, if $\mathcal{A}_{n,m}$ is the set of all alignments of
a sequence of length $n$ to a sequence of length $m$, and
$h \in \mathcal{A}_{n,m}$, then $h$ can be represented by a 
path in the {\em alignment graph}, or a sequence
of pairs of the form $(\sigma^1_i \diamond \sigma^2_j)$, $(\sigma^1 \diamond -)$,
or $(\sigma^2 \diamond -)$ where the symbol $\diamond$ is used to 
indicate the alignment of two characters. 
For $h \in \mathcal{A}_{n,m}$ let
\begin{itemize}
\item $h_H$ = $\{(i,j): (\sigma^1_i \diamond \sigma^2_j) \in h\}$,
\item $h_D$ = $\{i : (\sigma^1_i \diamond -) \in h\}$, 
\item $h_I$ = $\{j : (\sigma^2_j \diamond -) \in h\}$.
\end{itemize}
Less formally, $h_H$ is the set of position pairs in $\sigma^1$ and $\sigma^2$ that are aligned according to $h$, 
$h_D$ is the set of position in $\sigma^1$ that are gapped, and
$h_D$ is the set of position in $\sigma^2$ that are gapped.
Note that for any $h \in \mathcal{A}_{n,m}$
\begin{equation}
  |h_H| + |h_D| = n \textrm{ and } |h_H| + |h_I| = m.
\end{equation}
Two alignments are equivalent if they align the same character pairs, while the order of insertions and
deletions between any two consecutive aligned pairs is redundant. We therefore define:
\begin{equation}
  h^i \sim h^j \textrm{ if and only if } h^i_H = h^j_H \forall h^i,h^j \in \mathcal{A}_{n,m}.
\end{equation}
Note that $h^i_H = h^j_H$ if and only if $h^i_I = h^j_I$ and $h^i_D =
h^j_D$. We can therefore use the following equivalent definition:
\begin{eqnarray}
  h^i \sim h^j \textrm{ if and only if } h^i_I = h^j_I \textrm{ and }  h^i_D = h^j_D \nonumber\\ \forall  h^i,h^j \in \mathcal{A}_{n,m}.
\end{eqnarray}
We say that two alignments are {\em distinct} if they are not
equivalent.
The number of distinct alignments is in bijection with lattice paths
in the square grid (proof omitted):
\begin{prop}
The number of distinct alignments in $\mathcal{A}_{n,m}$ is ${n+m \choose m}$.
\label{prop:alcard}
\end{prop}
Given a predicted alignment $h^p$ and a reference alignment $h^r$, the $f_D$ and $f_M$ measures are defined:
\begin{eqnarray}
  f(h^i,h^j) = \frac{|h^i_H \cap h^j_H|}{|h^i_H|}\label{eqn:f}\\
  f_D(h^p,h^r) = f(h^r,h^p) = \frac{|h^r_H \cap h^p_H|}{|h^r_H|}\\
  f_M(h^p,h^r) = f(h^p,h^r) = \frac{|h^r_H \cap h^p_H|}{|h^p_H|}
\end{eqnarray}
Note that both measures do not explicitly use the $I$ and $D$ characters in $h^r$
and $h^p$, and are not well defined when $h^r$
or $h^p$ do not include any $H$ characters.

A distance function between any two alignments
$h^i,h^j \in \mathcal{A}_{n,m}$ is needed
in order to evaluate the quality of a predicted alignment given a reference alignment.
Such a distance function should satisfy:
\begin{align}
  & d(h^i,h^j) \ge 0 && \forall h^i,h^j \in \mathcal{A}_{n,m},\label{req1}\\
  & d(h^i,h^j) = 0 \textrm{ iff } h^i \sim h^j && \forall h^i,h^j \in \mathcal{A}_{n,m},\label{req2}\\
  & d(h^i,h^j) = d(h^j,h^i) && \forall h^i,h^j \in \mathcal{A}_{n,m},\label{req3}\\
  & d(h^i,h^j) + d(h^j,h^k) \ge d(h^i,h^k) && \forall h^i,h^j,h^k \in \mathcal{A}_{n,m}\label{req4}.
\end{align}
While these requirements are not satisfied by \eqref{eqn:f},
they are satisfied by the following:
\begin{eqnarray}
d(h^i,h^j) & = & 2 |h^i_H| + |h^i_I| + |h^i_D| - 2 |h^i_H \cap h^j_H| \nonumber\\ 
           &   & - |h^i_I \cap h^j_I| - |h^i_D \cap h^j_D|\nonumber\\
           & = & 2 |h^j_H| + |h^j_I| + |h^j_D| - 2 |h^i_H \cap h^j_H| \nonumber\\
           &   & - |h^i_I \cap h^j_I| - |h^i_D \cap h^j_D|\nonumber\\
           & = & n + m - 2 |h^i_H \cap h^j_H| \nonumber\\ 
           &   & - |h^i_I \cap h^j_I| - |h^i_D \cap h^j_D|).
\end{eqnarray}

\begin{prop}
$d(h^i,h^j)$ is a finite metric for $\mathcal{A}_{n,m}$.
\end{prop}
\emph{Proof}: It is easy to see that $d(h^i,h^j)$ satisfies requirements \eqref{req1}, \eqref{req2}, and \eqref{req3}.
We need to show that it satisfies the triangle inequality \eqref{req4}.
Let $U_{ij} = 2 |h^i_H \cap h^j_H| + |h^i_I \cap h^j_I| + |h^i_D \cap h^j_D|$, and 
$U_{ijk} = 2 |h^i_H \cap h^j_H \cap h^k_H| + |h^i_I \cap h^j_I \cap h^k_I| + |h^i_D \cap h^j_D \cap h^k_D|$.
Using the fact that $U_{ik} - U_{ijk} \ge 0$ and $U_{ij} + U_{jk} - U_{ijk} \le n + m$, we have that 
$d(h^i,h^j) + d(h^j,h^k) - d(h^i,h^k) = n + m - U_{ij} - U_{jk} + U_{ik}$
$= n + m  - (U_{ij} + U_{jk} - U_{ijk}) + U_{ik} - U_{ijk} \ge 0$. 

\begin{ex}[Metric for $\mathcal{A}_{2,2}$]
By Proposition \ref{prop:alcard}, there are six distinct alignments in $\mathcal{A}_{2,2}$. 
The metric is:
\end{ex}
{\normalsize
\begin{center}
\begin{tabular}{l | c @{ } c @{ } c @{ } c @{ } c @{ } c}
       & $HH$ & $HDI$ & $DIH$ & $IHD$ & $DHI$ & $DDII$\\
\hline
$HH$   & 0    & 2     & 2     & 4     & 4     & 4     \\
$HDI$  & 2    & 0     & 4     & 3     & 3     & 2     \\
$DIH$  & 2    & 4     & 0     & 3     & 3     & 2     \\
$IHD$  & 4    & 3     & 3     & 0     & 4     & 2     \\
$DHI$  & 4    & 3     & 3     & 4     & 0     & 2     \\
$DDII$ & 4    & 2     & 2     & 2     & 2     & 0     
\end{tabular}
\end{center}
}

Intuitively, the distance between two alignments is the total number of characters from both sequences that are aligned differently
in the two alignments.
Alternatively, the quantity $g(h^i,h^j) = 1 - \frac{d(h^i,h^j)}{n + m}$ is a convenient similarity measure that
can be interpreted as the fraction of characters that are aligned the same in both alignments.
We therefore define the \emph{Alignment Metric Accuracy (AMA)} of a predicted alignment $h^p$ given a reference alignment
$h^r$ to be $g(h^p,h^r)$. The intuitive motivation for this accuracy measure is that it represents the
fraction of characters in $\sigma^1$ and $\sigma^2$ that are correctly aligned, either to another character
or to a gap.

AMA can easily be extended to multiple sequence alignments (MSA) by using the sum-of-pairs approach.
Let $\mathcal{A}_{n_1,n_2,\ldots,n_k}$ be the space of all MSAs of $k$ sequences of lengths $n_1$ to $n_k$.
Given two MSAs $h^i,h^j \in \mathcal{A}_{n_1,n_2,\ldots,n_k}$, 
\begin{equation}
  d(h^i,h^j) = \sum_{s^1 = 1}^{k - 1} \sum_{s^2 > s^1}^k d(h^i_{s^1,s^2},h^j_{s^1,s^2}),
\end{equation}
where $h^i_{s^1,s^2}$ is the pairwise alignment of sequences $s^1,s^2$ as projected from the
MSA $h^i$ with all-gap columns removed. The similarity of two MSAa is defined to be
\begin{equation}
  g(h^p,h^r) = 1 - \frac{d(h^p,h^r)}{(k-1)\sum_{i=1}^k n_i}.
\end{equation}
Unlike standard sum-of-pairs scoring, our definition follows from 
the requirement that our accuracy measure should be based on a metric, and
the multiple AMA retains the desirable properties of the pairwise AMA.

\section{AMA Based Alignments}
\label{sec:meama}
\subsection{Maximum expected accuracy alignments}
Given a probabilistic model for alignments, such as a pair-HMM,
an alignment of a pair of sequences is typically obtained by the Viterbi 
algorithm \cite{viterbi67}, which finds the global alignment with highest probability.
In the case of a pair-HMM with three states, the Viterbi algorithm is equivalent
to the standard Needleman-Wunsch algorithm with affine gap scores \cite{durbin98}.
In effect, the Viterbi algorithm maximizes the expected number of times that a predicted
alignment is identical to the reference alignment ($h^p = h^r$). However, when the probability
of the most likely alignment is low it might be more desirable to predict alignments that 
are likely to align the most number of characters correctly on average
even if they are less likely to be identical to the correct alignment.

An alternative to Viterbi alignment is the \emph{optimal accuracy} alignment \cite{durbin98},
also called \emph{maximum expected accuracy} (MEA) alignment \cite{do05}, which maximizes the
expected $f_D$ score. The MEA alignment is calculated
using a dynamic programming algorithm that finds the alignment $h^p$ that maximizes
the expected number of correctly aligned character pairs:
\begin{equation}\label{eqn:oldaccuracy}
  h^p = \argmax_{h \in \mathcal{A}_{n,m}} \sum_{(i,j) \in h_H} \mathrm{P}(\sigma^1_i \Diamond \sigma^2_j | \sigma^1,\sigma^2,\theta),
\end{equation}
where $\mathrm{P}(\sigma^1_i \Diamond \sigma^2_j | \sigma^1,\sigma^2,\theta)$ is the posterior probability that
$\sigma^1_i$ is homologous to $\sigma^2_j$ given $\sigma^1$, $\sigma^2$ and the parameters of the model $\theta$.
In the case of a pair-HMM, these posterior probabilities can be computed in $O(n m)$ time
using the \emph{Forward-Backward} algorithm \cite{durbin98}.

\subsection{The AMAP algorithm}
\label{sec:amap}
While the MEA algorithm maximizes the expected $f_D$ score it can perform very poorly on the
$f_M$ score when the reference alignment contains many unaligned characters (gaps), since it tends to over-align characters.
Maximizing the expected $f_M$ score can be done easily by only aligning the pair
of characters with highest posterior probability to be homologous.
This will result in an alignment with only one $H$ character, $n-1$ $D$ characters, and $m-1$ $I$ characters, which
in most cases will result in a poor $f_D$ score. There is currently no alignment algorithm that allows for the adjustment of the
sensitivity/specificity tradeoff ($f_D$ / $f_M$ tradeoff). However, we
show that it is possible to maximize the expected AMA value using an
algorithm similar to the original MEA algorithm. By maximizing the
expected AMA, we avoid the problems of MEA alignment: in contrast to
the $f$ function, the $g$ function is symmetric, and therefore the sensitivity ($g(h^r,h^p)$) equals the
specificity ($g(h^p,h^r)$). In addition to maximizing the expected
AMA value, the new algorithm, which we call AMAP allows for the modification of one
parameter, which we term \emph{gap-factor},or $G_f$, to tune the $f_D$/$f_M$ tradeoff.

Let $\mathrm{P}(\sigma^1_i \Diamond - | \sigma^1,\sigma^2,\theta)$ be the posterior probability that $\sigma^1_i$ is not
homologous to any character in $\sigma^2$, and $\mathrm{P}(\sigma^2_j \Diamond - | \sigma^1,\sigma^2,\theta)$ the posterior
probability that $\sigma^2_j$ is not homologous to any character in $\sigma^1$. AMAP
should find the alignment $h^p$ that maximizes the expected number of characters that are correctly aligned to
another character or to a gap:
\begin{equation}\label{eqn:newaccuracy}
  \begin{split}
    h^p = \argmax_{h \in \mathcal{A}_{n,m}} \sum_{(i,j) \in h^p_H} \mathrm{P}(\sigma^1_i \Diamond \sigma^2_j | \sigma^1,\sigma^2,\theta) &+ \\
    G_f \sum_{i \in h^p_D} \mathrm{P}(\sigma^1_i \Diamond - | \sigma^1,\sigma^2,\theta) &+ \\
    G_f \sum_{j \in h^p_I} \mathrm{P}(\sigma^2_j \Diamond - | \sigma^1,\sigma^2,\theta).
  \end{split}
\end{equation}

Note that when $G_f = 0.5$ the algorithm maximizes the expected AMA value, 
while when $G_f = 0$ the expression in \eqref{eqn:newaccuracy}
is equal to the expression in \eqref{eqn:oldaccuracy},
and the algorithm is identical to the original MEA algorithm, which maximized the expected $f_D$ score.
Setting $G_f$ to higher values than $0.5$ results in better $f_M$ scores in the expense of lower $f_D$ scores.
In effect, the gap-factor parameter allows for the tuning of the $f_D$/$f_M$ tradeoff.

An MEA subroutine can be used to construct multiple alignments using a number of different 
strategies, one of which is progressive alignment. Expected AMA maximization
was performed in this context by 
using the ProbCons platform (the code is available in open source under the Gnu public license)
with the MEA algorithm modified so that the developer score is no longer maximized.
Our resulting AMAP algorithm also omits the heuristic consistency
transformation of ProbCons, and simply aligns pairwise alignments along a guide tree. 

\section{Results}
\label{sec:results}
\begin{table*}[!tb]
\begin{center}
\small{
\begin{tabular}{l c c c c c c c c c c c c}
        & \multicolumn{3}{c}{Twilight} & \multicolumn{3}{c}{Superfamilies} &  \multicolumn{3}{c}{Twilight-FP} & \multicolumn{3}{c}{Superfamilies-FP} \\
\cline{2-13}
Program & $f_D$ & $f_M$ & $AMA$ & $f_D$ & $f_M$ & $AMA$ & $f_D$ & $f_M$ & $AMA$ & $f_D$ & $f_M$ & $AMA$\\
\hline
Align-m & 21.6 & \textbf{23.6} & \textbf{51.7} & 49.2 & \textbf{45.6} & \textbf{56.9} & 17.8 & \textbf{6.4} & \textbf{81.5} & 44.8 & \textbf{16.8} & \textbf{77.5}\\
CLUSTALW & 25.6 & 14.7 & 24.9 & 54.0 & 38.1 & 43.8 & 20.4 & 2.4 & 35.5 & 50.9 & 7.4 & 37.0\\
MUSCLE & 27.3 & 16.4 & 27.6 & 56.3 & 40.3 & 46.4 & 19.4 & 2.3 & 37.1 & 49.7 & 7.5 & 38.9\\
ProbCons & \textbf{32.1} & 21.1 & 37.3 & \textbf{59.8} & 44.4 & 51.8 & \textbf{26.7} & 4.4 & 55.7 & 56.0 & 10.9 & 55.0\\
T-Coffee & 29.4 & 19.6 & 35.6 & 58.4 & 43.7 & 50.9 & 26.5 & 4.2 & 54.1 & \textbf{57.0} & 11.0 & 54.4
\end{tabular}
}
\caption{\textbf{Performance of aligners on the SABmark benchmark datasets.} Entries show the average developer ($f_D$), 
modeler ($f_M$) and alignment metric accuracy ($AMA$). Best results are shown in bold. All numbers have been multiplied by 100.}
\label{tbl:SABmarkComp}
\end{center}
\end{table*}
\subsection{Performance of existing programs on the SABmark datasets}
We began by assessing the performance of
existing programs on the SABmark 1.65 \cite{vanWalle05} datasets 
with the goal of comparing alignment metric accuracy
with previously used measures. SABmark includes two sets of pairwise reference alignments
with known structure from the ASTRAL \cite{brenner00} database. The Twilight Zone set
contains 1740 sequences with less than 25\% identity divided into 209 groups based on SCOP folds \cite{Murzin95}.
The Superfamilies set contains 3280 sequences with less than 50\% identity divided into 425 groups.
Additionally, each dataset has a ``false positives'' version, which contains unrelated sequences 
with the same degree of sequence similarity in addition to the related sequences.

Table \ref{tbl:SABmarkComp} shows the performance of a number of existing alignment programs
as measured by the developer, modeler, and AMA accuracy measures
on the four SABmark 1.65 datasets. Methods tested include
Align-m 2.3 \cite{vanWalle05}, CLUSTALW 1.83 \cite{thompson94}, MUSCLE 3.52 \cite{edgar04}, 
ProbCons 1.1 \cite{do05} and T-Coffee 2.49 \cite{notredame00}.
The results highlight the inherent sensitivity/specificity tradeoff.
While ProbCons and T-Coffee have the best developer scores, Align-m has the best modeler scores.
It is not clear which program outperforms the others. Programs with higher sensitivity tend to over-align
unalignable regions, which results in lower specificity. We would like to answer the question, 
which program produces alignments that are the closest to the reference alignments? This is exactly the
interpretation of the new AMA measure. Using this measure it is clear that Align-m is the most accurate
alignment program among the ones tested on the SABmark benchmark
datasets.\footnote{Note that the SABmark benchmark dataset was
compiled by the same authors as Align-m.} 

\subsection{Controls for multiple alignment experiments}
\begin{table*}[!tb]
\begin{center}
\footnotesize{
\begin{tabular}{l c c c c c c c c}
         & Align-m & CLUSTALW & MUSCLE & ProbCons & T-Coffee & AMAP & AMAP-4 & Reference\\
Align-m  & & 37.3 & 39.9 & 52.8 & 50.4 & 64.7 & 68.7 & \textbf{67.0}\\
CLUSTALW & 45.0 & & 38.2 & 38.1 & 39.1 & 39.5 & 35.9 & \textbf{37.0}\\
MUSCLE   & 43.8 & 49.4 & & 43.2 & 42.3 & 43.9 & 38.9 & \textbf{39.2}\\
ProbCons & 32.0 & 48.7 & 47.1 & & 52.0 & 67.2 & 53.4 & \textbf{51.1}\\
T-Coffee & 30.1 & 47.0 & 45.9 & 38.7 & & 56.6 & 50.9 & \textbf{50.1}\\
AMAP     & 15.5 & 45.1 & 43.7 & 29.8 & 29.4 & & 72.9 & \textbf{64.4}\\
AMAP-4 & 13.3 & 45.6 & 44.3 & 32.1 & 29.9 & 11.6 & & \textbf{70.2}\\
Reference & \textbf{13.7} & \textbf{44.1} & \textbf{43.0} & \textbf{31.1} & \textbf{28.6} & \textbf{13.8} & \textbf{11.4} &
\end{tabular}
}
\caption{\textbf{Total distance and average similarity ($g$) of different aligners on the SABmark dataset.} 
Values below the diagonal show the total distance between alignments produced by different alignment programs.
Values above the diagonal show the average similarity ($g$) between the different alignments. 
Distance values have been divided by one million, and similarity values have been multiplied by 100.
AMAP-4 is the AMAP algorithm with gap-factor of 4.}
\label{tbl:distances}
\end{center}
\end{table*}

The reference alignments in SABmark are themselves somewhat
subjective, as they are based on structural alignment programs.
A recurring question has been how
to judge the accuracy of alignment in the absence of a reference. To
demonstrate how AMA is useful for that, we compared
the total distance and average similarity ($g$) between the alignments produced by the 
five alignment programs, two variants of the AMAP algorithm, and the reference alignments.
Table \ref{tbl:distances} shows these values averaged over the entire SABmark dataset.
An interesting observation is that there is a correlation between the distances of 
the alignments of any given program to the reference alignment, and their distances to the 
closest alignments of other programs. For example, CLUSTALW alignments are $37\%$ similar 
on average to the reference SABmark alignments, and $39.5\%$ similar to AMAP alignments, 
while Align-m alignments are $67\%$ similar to the reference alignments, and $68.7\%$ similar
to AMAP-4 alignments.

\begin{figure}[!tb]
  \begin{center}
  \includegraphics[angle=270,width=0.5\textwidth]{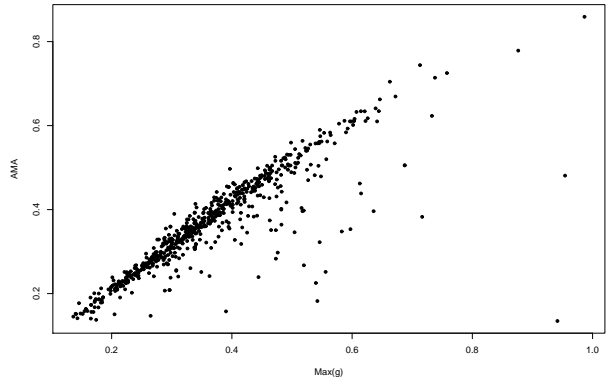}
  \end{center}
\caption{\textbf{Correlation between the AMA of CLUSTALW (as judged by reference
alignments in SABmark), and distance to the nearest alignment produced by another program.}
Each dot in the plot corresponds to one CLUSTALW alignment in the
SABmark Twilight-FP and Superfamilies-FP datasets. The $x$ coordinate
represents the similarity ($g$) of the CLUSTALW alignment to the
closest alignment produced by one of four other programs (Align-m,
MUSCLE, ProbCons, T-Coffee). The $y$
coordinate represents the Alignment Metric Accuracy (AMA) of the
CLUSTALW alignment as judged by the reference SABmark alignment.}
\label{fig:correlation}
\end{figure}

We propose to use the similarity of predicted alignments from
different alignment program as a control before using a predicted
alignment. Figure \ref{fig:correlation} shows the correlation between
the accuracy (AMA) of of CLUSTALW alignments of the Twilight-FP and
Superfamilies-FP datasets, and their maximum similarity to any of the
alignments produced by the other four programs.  It is evident that
there is a strong correlation between the two values. We observed
similar correlation for the other alignment programs (see
supplementary data). In the case of the two datasets with no false
positives (Twilight and Superfamilies) the maximum similarity behaves
more as an upper bound rather than a strong predictor for the AMA value (data not shown). This is still useful for discarding alignments that are not similar to other predicted alignments,
since they are very unlikely to be similar to the true alignment. 

To summarize, the following protocol can be used to identify bad alignments
in the absence of a reference alignment: 
\begin{enumerate}
\item Align the target sequences with a preferred alignment program.
\item Align the target sequences with all other available alignment programs.
\item Measure the similarity of the first alignment to every other alignment,
\item If the similarity of the closest alignment to the first alignment is below a certain threshold, discard the alignment.
\end{enumerate}
The above procedure has no mathematical guarantees, but our empirical results show that most of the discarded alignments will have
an AMA value less than the selected threshold.

\subsection{Performance of the AMAP algorithm}
\begin{table*}[!tb]
\begin{center}
\begin{tabular}{l c c c}
\hline
\multicolumn{4}{c}{Default transition probabilities}\\
Algorithm & $f_D$ & $f_M$ & AMA\\
\hline
Viterbi & 27.2 & 16.3 & 28.0\\
$G_f=0$ & \textbf{29.6} & 17.7 & 29.0\\
$G_f=0.5$ & 28.1 & 19.7 & 37.4\\
$G_f=1$ & 25.8 & 22.8 & 45.2\\
$G_f=2$ & 22.4 & 27.6 & 51.5\\
$G_f=4$ & 18.9 & 33.1 & 54.8\\
$G_f=8$ & 15.9 & 38.9 & 56.4\\
$G_f=12$ & 14.3 & 43.3 & 56.7\\
$G_f=16$ & 13.1 & 46.1 & 56.8\\
$G_f=20$ & 12.4 & 48.0 & \textbf{56.8}\\
$G_f=28$ & 11.3 & \textbf{51.5} & 56.7\\
\hline
\end{tabular}
\begin{tabular}{l c c c}
\hline
\multicolumn{4}{c}{``Correct'' transition probabilities}\\
Algorithm & $f_D$ & $f_M$ & AMA\\
\hline
Viterbi   & 18.8 & 17.0 & 46.7\\
$G_f=0$   & \textbf{25.9} & 17.5 & 37.2\\
$G_f=0.5$ & 17.2 & 34.7 & 56.3\\
$G_f=1$   & 14.1 & 42.2 & \textbf{57.3}\\
$G_f=2$   & 11.3 & 52.2 & 57.3\\
$G_f=4$   & 8.9 & 59.3 & 56.7\\
$G_f=8$   & 7.0 & 68.7 & 56.0\\
$G_f=12$   & 6.1 & 74.5 & 55.6\\
$G_f=16$   & 5.5 & 77.3 & 55.4\\
$G_f=20$   & 5.1 & 80.2 & 55.2\\
$G_f=28$   & 4.6 & \textbf{83.1} & 55.0\\
\hline
\end{tabular}
\caption{\textbf{Performance of algorithm variants on the SABmark Twilight Zone set.}
Entries show the $f_D$, $f_M$, and AMA scores of the Viterbi, and
AMAP alignments with different gap-factor ($G_f$) values on the
SABmark Twilight Zone set, which includes 209 alignment groups.
The first five columns show the results
using default transition probabilities, and the last five columns show the results using
transition probabilities calculated for each group from the reference alignments.
All scores have been averaged over groups and multiplied by 100.}
\label{tbl:twilight}
\end{center}
\end{table*}

\begin{table*}[!tb]
\begin{center}
\begin{tabular}{l c c c}
\hline
\multicolumn{4}{c}{Default transition probabilities}\\
Algorithm & $f_D$ & $f_M$ & AMA\\
\hline
Viterbi & 53.1 & 38.1 & 44.2\\
$G_f=0$ & \textbf{54.8} & 39.3 & 45.2\\
$G_f=0.5$ & 53.6 & 42.0 & 49.8\\
$G_f=1$ & 51.6 & 46.2 & 54.5\\
$G_f=2$ & 48.1 & 52.1 & 58.2\\
$G_f=4$ & 44.1 & 58.5 & 59.9\\
$G_f=6$ & 41.8 & 62.0 & \textbf{60.2}\\
$G_f=8$ & 40.2 & \textbf{64.3} & 60.1\\
\hline
\end{tabular}
\begin{tabular}{l c c c}
\hline
\multicolumn{4}{c}{``Correct'' transition probabilities}\\
Algorithm & $f_D$ & $f_M$ & AMA\\
\hline
Viterbi & 46.8 & 41.3 & 53.3\\
$G_f=0$ & \textbf{52.4} & 40.1 & 49.2\\
$G_f=0.5$ & 45.4 & 56.1 & 60.5\\
$G_f=1$ & 41.8 & 63.4 & \textbf{61.5}\\
$G_f=2$ & 37.9 & 70.9 & 61.2\\
$G_f=4$ & 34.1 & 75.9 & 60.0\\
$G_f=6$ & 31.9 & 78.4 & 59.2\\
$G_f=8$ & 30.5 & \textbf{79.9} & 58.6\\
\hline
\end{tabular}
\caption{\textbf{Performance of algorithm variants on the SABmark Superfamilies set.}
Entries show the $f_D$, $f_M$, and AMA scores of the Viterbi, and
AMAP alignments with different gap-factor ($G_f$) values on the
SABmark Superfamilies set, which includes 425 alignment groups.
The first five columns show the results
using default transition probabilities, and the last five columns show the results using
transition probabilities calculated for each group from the reference alignments.
All scores have been averaged over groups and multiplied by 100.
}
\label{tbl:superfamilies}
\end{center}
\end{table*}

\begin{table*}[!tb]
\begin{center}
\begin{tabular}{@{} r @{~~} r @{~~} r @{~~}|| r @{~~} r @{~~} r @{~~} r r @{~~} r @{~~} r @{~~} r r @{~~} r @{~~} r @{~~} r @{}}
\hline
\multicolumn{3}{c}{Parameter Settings} & \multicolumn{4}{c}{$f_D$} & \multicolumn{4}{c}{$f_M$} & \multicolumn{4}{c}{AMA}\\
$e_{match}$ & $\delta$ & $\varepsilon$ & 0 & 0.5 & 1 & Vit & 0 & 0.5 & 1 & Vit & 0 & 0.5 & 1 & Vit\\
\hline
30 & 10.0 & 90 & \textbf{7.1} & 1.1 & 0.8 & 0.6 & 4.0 & 61.7 & \textbf{73.7} & 2.2 & 9.5 & \textbf{51.0} & 50.9 & 44.7\\
50 & 10.0 & 90 & \textbf{23.6} & 3.9 & 2.1 & 12.5 & 15.7 & 77.0 & \textbf{83.5} & 10.8 & 27.2 & \textbf{52.2} & 51.4 & 30.3\\
60 & 10.0 & 90 & \textbf{33.6} & 17.7 & 12.2 & 24.6 & 23.2 & 69.5 & \textbf{82.9} & 23.8 & 34.5 & \textbf{57.2} & 55.9 & 41.2\\
80 & 10.0 & 90 & \textbf{61.2} & 53.4 & 46.6 & 53.6 & 45.9 & 72.1 & \textbf{81.9} & 51.7 & 57.3 & \textbf{70.8} & 70.7 & 62.4\\
80 & 5.0 & 90 & \textbf{83.2} & 80.8 & 77.5 & 81.1 & 76.3 & 85.3 & \textbf{89.9} & 77.7 & 78.8 & \textbf{82.4} & 82.2 & 78.6\\
80 & 2.0 & 98 & \textbf{94.4} & 93.5 & 92.7 & 92.9 & 89.0 & 95.0 & \textbf{96.2} & 93.6 & 93.0 & \textbf{95.4} & 95.3 & 94.6\\
80 & 0.9 & 98 & \textbf{97.4} & 97.2 & 96.6 & 96.8 & 95.2 & 97.5 & \textbf{98.2} & 96.6 & 96.2 & \textbf{97.2} & \textbf{97.2} & 96.7\\
30 & 0.1 & 98 & \textbf{94.5} & \textbf{94.5} & 93.8 & 90.1 & 93.7 & 93.8 & \textbf{94.1} & 89.3 & \textbf{93.1} & \textbf{93.1} & 92.6 & 88.5\\
50 & 0.1 & 98 & \textbf{99.4} & \textbf{99.4} & \textbf{99.4} & 99.3 & \textbf{99.4} & \textbf{99.4} & \textbf{99.4} & 99.3 & \textbf{99.2} & \textbf{99.2} & \textbf{99.2} & 99.0\\
\multicolumn{3}{c||}{unrelated}& 100 & 100 & 100 & 100 & 0.0 & 0.0 & 0.0 & 0.0 & 72.2 & 96.8 & \textbf{98.4} & 94.3\\
\hline
\end{tabular}
\caption{\textbf{Performance of algorithm variants on simulated data.}
Entries show the performance of the Viterbi algorithm (Vit), and the AMAP algorithm with
different settings of the gap-factor parameter (0, 0.5, and 1) using three accuracy measures ($f_D$, $f_M$, and AMA).
The first three columns show the configuration of the pair-HMM parameters $e_{match}$
(match emission probability), $\delta$ (gap initiation probability) and $\varepsilon$
(gap extension probability),
except for the last row for which random unrelated sequences have been aligned.
Best results for every parameter configuration and measure are shown in bold.
All numbers have been multiplied by 100.}
\label{tbl:sim}
\end{center}
\end{table*}

Next, we investigated, using pairwise alignments, whether the AMAP algorithm
can improve on the Viterbi and the Maximum Expected Accuracy (MEA) alignment algorithms 
for maximizing the AMA.

We first evaluated the algorithms with the same default parameters
used in ProbCons ($\delta = 0.01993$, $\varepsilon = 0.79433$,
$\pi_{match} = 0.60803$, and emission probabilities based on the
BLOSUM62 matrix). Table \ref{tbl:twilight} shows the results of the
Viterbi algorithm and the AMAP algorithm with different gap-factor
values on the SABmark Twilight Zone set, and table
\ref{tbl:superfamilies} show the results on the SABmark Superfamilies
set.

The results on both sets show the expected correlation between the gap-factor value and the $f_M$ score,
and the negative correlation between the gap-factor value and the $f_D$ score. This validates the prediction that
the gap-factor can be used as a tuning parameter for the sensitivity/specificity tradeoff of matched characters.

When the gap-factor is set to $0.5$ or higher the alignment accuracy is significantly better than Viterbi alignments,
when the original MEA algorithm (AMAP with gap-factor set to 0) is used, the alignment accuracy is almost identical to that of the Viterbi algorithm.
The most accurate alignments were achieved by setting the gap-factor to values higher than $0.5$
(20 in the Twilight Zone set and 6 in the Superfamilies set). We suspected that this is due to the fact 
that the default pair-HMM parameters underestimate the probability of insertion and deletions. 
To validate that we calculated the ``true'' transition probabilities for
each alignment group using the reference alignments, and repeated the experiment.

The performance of the algorithms using the ``correct''
transition probabilities are shown in the right columns 
of tables \ref{tbl:superfamilies} and \ref{tbl:twilight}.
As expected, with the correct parameters, the accuracy of the alignments achieved when the gap-factor 
is set to $0.5$ are very close to the
best ($G_f=1$). Note that we did not modify the emission probabilities, which might be the 
reason $G_f=0.5$ did not maximize the actual accuracy.
These results show that the AMAP algorithm significantly outperforms the Viterbi and MEA algorithms 
(61.5 AMA compared to 53.3 and 49.2 AMA on the Superfamilies dataset, and 57.3 AMA compared to 46.7 and 37.2 
on the Twilight Zone dataset). 
Moreover, with the adjusted parameters, the Viterbi algorithm outperforms the MEA algorithm ($G_f=0$) on both datasets.
This is due to the over-alignment problem of the MEA algorithm, which uses the expected $f_D$ score 
as its objective function at the
expense of the $f_M$ and AMA scores. 
Note that the best AMA scores achieved with the default transition probabilities
are very close to those of the correct probabilities, demonstrating that adjustment of
the gap-factor parameter is able to compensate for bad estimation of the parameters 
of the underlying probabilistic model.

\begin{table*}[!tb]
\begin{center}
\normalsize{
\begin{tabular}{l c c c c c}
Program & Twilight & Superfamilies & Twilight-FP & Superfamilies-FP & Overall\\
\cline{2-6}
Align-m & 51.7 & 56.9 & 81.5 & 77.5 & 67.0\\
ProbCons & 37.3 & 51.8 & 55.7 & 55.0 & 51.1\\
AMAP & 46.1 & 56.3 & 75.8 & 75.8 & 64.4\\
AMAP-4 & \textbf{52.2} & \textbf{57.9} & \textbf{84.2} & \textbf{84.6} & \textbf{70.2}
\end{tabular}
}
\caption{\textbf{Performance of selected programs on the SABmark benchmark datasets.} 
  Entries show the AMA score for each program and data set. All numbers have been multiplied by 100.}
\label{tbl:SABmarkResults}
\end{center}
\end{table*}

In order to further analyze the performance of the AMAP algorithm compared to the Viterbi and MEA algorithms, we also
conducted simulation studies. Table \ref{tbl:sim} compares the performance of the Viterbi, MEA, and AMAP variants
on different sets of simulated pairs of related and unrelated DNA
sequences.

Data was simulated using a pair-HMM to generate aligned pairs of nucleotide
sequences. The pair-HMM parameters included the
transition probabilities $\delta$ (gap initiation)
and $\varepsilon$ (gap extension).
For simplicity we fixed the initial probability $\pi_{match}$ of starting in a Match state to be
$1 - 2\delta$.
For the emission probabilities we used a simple
model that assigns equal probability ($\frac{1}{4}$) to any nucleotide in the Insert
or Delete states, $\frac{e_{match}}{4}$ probability for a match in the Match state, and
$\frac{1-e_{match}}{12}$ probability for a mismatch in the Match state, where $e_{match}$
is the probability to emit a pair of identical characters in the Match state.

For every setting of the parameters we generated 10 reference alignments with $\min(n,m)=1000$.
An identical pair-HMM with the same parameters was then used to compare the performance of the
Viterbi algorithm and MEA algorithm with gap-factor values of (0, 0.5, and 1).
We treat every set of 10 alignments as one big alignment, and calculate the accuracy ($g(h^p,h^r)$)
of the predicted alignments, the $f_D$, and $f_M$ scores.
In addition to the simulated reference alignment generated from the pair-HMM, we also
generated 10 pairs of unrelated random sequences of length 1000 each with equal probability
for every character. All algorithms have been evaluated on the resulting reference alignments, which
include no $H$ characters, using the probabilities 0.8, 0.1, and 0.9 for the $e_{match}$, $\delta$,
and $\varepsilon$ parameters respectively.

The simulation results demonstrate that the AMAP algorithm produces
alignments that are more accurate than the Viterbi and MEA alignment algorithms
on both closely related and distant sequences.
As expected the best $f_D$ scores are achieved using the MEA algorithm ($G_f = 0$), 
the best $f_M$ scores when $G_f = 1$, and the best AMA when $G_f = 0.5$.

It is interesting to note that for distant sequences with larger gap
initiation probability ($\delta$),  
the Viterbi algorithm has better AMA score than the
MEA algorithm ($G_f = 0$). This again emphasizes the main weakness of MEA, which tends to
over-align unalignable regions. 
This problem is even more pronounced when aligning unrelated sequences. 
The MEA algorithm performs poorly compared to the AMAP algorithm
and even the Viterbi algorithm, achieving a mere 72.2 AMA scored compared to 96.8 and 94.8 respectively. 
This is due to the fact that the MEA algorithm wrongly aligns 2781 character pairs, compared to
157, 316, and 572 in the case of $G_f = 1$, $G_f = 0.5$, and Viterbi alignment respectively (data not shown).

Finally, we compared the performance of the multiple sequence alignments version of the AMAP algorithm 
compared to ProbCons and Align-m. Table \ref{tbl:SABmarkResults} shows the AMA scores of each program
on the four SABmark datasets. AMAP and Align-m are clearly superior to ProbCons. While AMAP with default
parameters achieves slightly lower AMA scores than Align-m, setting the gap-factor to 4 produces the
most accurate alignments. This demonstrate the power of the AMAP algorithm, which can easily be tuned 
using a single parameter to improve alignment accuracy even when the parameters of the underlying 
statistical model (transition and emission probabilities of the Pair-HMM) do not fit the data very well.

\section{Discussion}
We have proposed a metric for the set of alignments, and shown how it can be used
both to judge the accuracy of alignments, and as the basis for an optimization criteria
for alignment. The importance of the metric lies in the fact that if two alignments
are far from each other, we can conclude that at least one of them is inaccurate. This is 
a direct consequence of the triangle inequality. More importantly, we show that when alignments made
by widely used software programs
are compared to each other they are far apart, thus quantitatively confirming that
multiple alignment is a difficult problem. Although we see that the sensitivity of many programs
is high, i.e., many of the residues that should be aligned together  are correctly aligned, it is 
also the case that many residues are incorrectly aligned. This is particularly
evident in results from the Twilight-FP and Superfamilies-FP datasets
that contain unrelated sequences. If functional inferences are to be
made from sequence alignments, it is therefore important to control
for specificity, and not only sensitivity. Our alignment algorithm,
AMAP, which maximizes the expected AMA, outperforms existing programs
on benchmark datasets.

Most exiting multiple alignment benchmark datasets include only
alignments of ``core blocks'', and it is therefore only possible to measure the
sensitivity of matches ($f_D$), and not their specificity ($f_M$) or
the AMA. However, the fact that it is harder to construct datasets that
allow for measuring the latter two does not mean that alignment algorithms
should maximize sensitivity at the expense of specificity. Our
AMAP algorithm is the first to allows the user to tune the inherent
sensitivity/specificity tradeoff using the gap-factor parameter.  In
many cases, such as when using MSA for phylogenetic tree
reconstruction, or for identification of remote homology, higher
specificity is preferred over higher sensitivity.  In addition, as we
have shown, tuning the gap-factor parameter can in some cases
compensate for poor parameter estimation of the underlying
probabilistic model (pair-HMM).  Further work is needed to develop
methods for automatic adjustment of this parameter for a given dataset
when the probabilistic parameters do not fit the data very well, and a
reference alignment is not available.

In the typical case 
where reference alignments are not available, our empirical
observation that the distances between alignments correlate strongly
with the accuracy of the programs that generated them can be used to
discard inaccurate alignments. It is possible that more sophisticated
strategies based on this principle could further help in
quantitatively assessing alignment reliability.

We have not discussed the relevance of our results to DNA multiple alignments, however 
many of our ideas can be easily adapted. As with protein multiple alignment, 
the focus in DNA alignment has been on sensitivity rather than specificity. For example, 
whole genome alignments are often judged by exon coverage. 
We have focused on protein sequence in this
initial study, because in certain cases reference alignments can be constructed based on structural
alignments.

Finally, we mention that it is possible to formulate MEA multiple alignment using our AMA as an 
integer linear program using ideas similar to those in \cite{althaus02,prestwich03}. In particular, 
it is possible to set up a program with a polynomial number of variables and constraints. 
It should be interesting to study the possibility of applying approximation methods to solving
such programs.

\section*{Acknowledgments}
A.S was partially supported by NSF grant EF 03-31494. 
G.M. was supported by the Max-Planck / Alexander von Humboldt International Research Prize.
L.P. was partially supported by a Sloan Research Fellowship. 
\bibliography{references}

\begin{thebibliography}{10}

\bibitem{althaus02}
Ernst Althaus, Alberto Caprara, Hans~Peter Lenhof, and Knut Reinert.
\newblock {Multiple sequence alignment with arbitrary gap costs: Computing an
  optimal solution using polyhedral combinatorics}.
\newblock {\em Bioinformatics}, 18(90002):4S--16, 2002.

\bibitem{batzoglou05}
S~Batzoglou.
\newblock The many faces of sequence alignment.
\newblock {\em Briefings in Bioinformatics}, 6:6--22, 2005.

\bibitem{blanchette04}
M~Blanchette, W~J Kent, C~Riemer, L~Elnitski, A~F~A Smit, K~M Roskin,
  R~Baertsch, K~Rosenbloom, H~Clawson, E~D Green, D~Haussler, and W~Miller.
\newblock Aligning multiple genome sequences with the threaded blockset
  aligner.
\newblock {\em Genome Research}, 14:708--715, 2004.

\bibitem{brenner00}
Steven~E. Brenner, Patrice Koehl, and Michael Levitt.
\newblock {The ASTRAL compendium for protein structure and sequence analysis}.
\newblock {\em Nucl. Acids Res.}, 28(1):254--256, 2000.

\bibitem{do05}
Chuong~B. Do, Mahathi~S.P. Mahabhashyam, Michael Brudno, and Serafim Batzoglou.
\newblock {ProbCons: Probabilistic consistency-based multiple sequence
  alignment}.
\newblock {\em Genome Res.}, 15(2):330--340, 2005.

\bibitem{durbin98}
R.~Durbin, S.~Eddy, A.~Krogh, and G.~Mitchison.
\newblock {\em Biological sequence analysis. Probablistic models of proteins
  and nucleic acids}.
\newblock Cambridge University Press, 1998.

\bibitem{edgar04}
Robert~C. Edgar.
\newblock {MUSCLE: multiple sequence alignment with high accuracy and high
  throughput}.
\newblock {\em Nucl. Acids Res.}, 32(5):1792--1797, 2004.

\bibitem{holmes98}
I.~Holmes and R.~Durbin.
\newblock Dynamic programming alignment accuracy.
\newblock {\em J. Comp. Biol.}, 5(3):493--504, 1998.

\bibitem{miller00}
W~Miller.
\newblock comparison of genomic sequences: Solved and unsolved problems.
\newblock {\em Bioinformatics}, 17:391--397, 2000.

\bibitem{Murzin95}
A.G. Murzin, S.E. Brenner, T.~Hubbard, and C.~Chothia.
\newblock Scop: a structural classification of proteins database for the
  investigation of sequences and structures.
\newblock {\em Journal of molecular biology}, 247(4):536, 1995.

\bibitem{notredame00}
C~Notredame, D~Higgins, and J~Heringa.
\newblock T-coffee: A novel method for multiple sequence alignments.
\newblock {\em Journal of Molecular Biology}, 302:205--217, 2000.

\bibitem{pachter05}
L.~Pachter and B.~Sturmfels, editors.
\newblock {\em Algebraic Statistics for Computational Biology}.
\newblock Cambridge University Press, 2005.

\bibitem{pecan}
B.~Paten.
\newblock http://www.ebi.ac.uk/$\sim$bjp/pecan/, 2005.

\bibitem{prestwich03}
S.~Prestwich, D.~Higgins, and O.~O'Sullivan.
\newblock Pseudo-boolean multiple sequence alignment.
\newblock In {\em Tenth Workshop on Automated Reasoning}, 2003.

\bibitem{sauder00}
J.M. Sauder, J.W. Arthur, and R.L. Dunbrack.
\newblock Large-scale comparison of protein sequence alignment algorithms with
  structure alignments.
\newblock {\em Proteins}, 40(1):6--22, 2000.

\bibitem{spiro04}
Peter~A. Spiro and Natasa Macura.
\newblock A local alignment metric for accelerating biosequence database
  search.
\newblock {\em Journal of Computational Biology}, 11(1):61--82, 2004.

\bibitem{sze05}
Sing-Hoi Sze, Yue Lu, and Qingwu Yang.
\newblock A polynomial time solvable formulation of multiple sequence
  alignment.
\newblock {\em Lecture Notes in Computer Science}, 3500:204--216, April 2005.

\bibitem{thompson94}
JD~Thompson, DG~Higgins, and TJ~Gibseon.
\newblock Clustalw: improving the sensitivity of progressive multiple sequence
  alignment through sequence weighting, position-specific gap penalties and
  weight matrix choice.
\newblock {\em Nucleic Acids Research}, 22:4673--4680, 1994.

\bibitem{vanWalle05}
Ivo Van~Walle, Ignace Lasters, and Lode Wyns.
\newblock {SABmark--a benchmark for sequence alignment that covers the entire
  known fold space}.
\newblock {\em Bioinformatics}, 21(7):1267--1268, 2005.

\bibitem{viterbi67}
Andrew~J. Viterbi.
\newblock Error bounds for convolutional codes and an asymptotically optimum
  decoding algorithm.
\newblock {\em IEEE Transactions on Information Theory}, IT-13(2):260--269,
  April 1967.

\end{thebibliography}
\end{document}